\documentclass[notitlepage,floatfix,aps,prl,reprint,amsmath,amssymb,twocolumn,superscriptaddress,10pt]{revtex4-2}
\usepackage{graphicx}% Include figure files
\usepackage{dcolumn}% Align table columns on decimal point
\usepackage{bm}
\usepackage{braket}
\usepackage{amssymb}
\usepackage{hyperref}
\usepackage{amsmath}
\usepackage{textcomp}
\usepackage{amsfonts}
\usepackage{gensymb}

\usepackage{xcolor}

\makeatletter
\let\oldabs\abs
\def\abs{\@ifstar{\oldabs}{\oldabs*}}

\begin{document}
\title{Coherence in cooperative photon emission from indistinguishable quantum emitters}

\author{Zhe Xian Koong}
\email[Correspondence: ]{zk49@hw.ac.uk}
\affiliation{
  SUPA, Institute of Photonics and Quantum Sciences, Heriot-Watt University, Edinburgh EH14 4AS, UK.
}
\author{Moritz Cygorek}
\author{Eleanor Scerri}
\affiliation{
  SUPA, Institute of Photonics and Quantum Sciences, Heriot-Watt University, Edinburgh EH14 4AS, UK.
}
\author{Ted  S. Santana}
\affiliation{
Centro de Ci\^{e}ncias Naturais e Humanas, Universidade Federal do ABC, Santo Andr\'{e}, São Paulo 09210-580, Brazil.
}
\author{Suk In Park}
\affiliation{
  Center for Opto-Electronic Materials and Devices Research, Korea Institute of Science and Technology, Seoul 02792, Republic of Korea.
}
\author{Jin Dong Song}
\affiliation{
  Center for Opto-Electronic Materials and Devices Research, Korea Institute of Science and Technology, Seoul 02792, Republic of Korea.
}
\author{Erik M. Gauger}
\author{Brian D. Gerardot}
\email[Correspondence: ]{b.d.gerardot@hw.ac.uk}
\affiliation{
  SUPA, Institute of Photonics and Quantum Sciences, Heriot-Watt University, Edinburgh EH14 4AS, UK.
}
\email{zk49@hw.ac.uk}

\date{March 18, 2022}
\begin{abstract}

  Photon-mediated interactions between atomic systems can arise via coupling to a common electromagnetic mode or by quantum interference.
  Here, we probe the role of coherence in cooperative emission arising from two distant but indistinguishable solid-state emitters because of path erasure. The primary signature of cooperative emission, the emergence of ``bunching" at zero delay in an intensity correlation experiment, is used to characterise the indistinguishability of the emitters, their dephasing, and the degree of correlation in the joint system that can be coherently controlled.
  In a stark departure from a pair of uncorrelated emitters, in Hong-Ou-Mandel type interference measurements we observe photon statistics from a pair of indistinguishable emitters resembling that of a weak coherent state from an attenuated laser.
  Our experiments establish techniques to control and characterize cooperative behavior between matter qubits using the full quantum optics toolbox, a key step toward realizing large-scale quantum photonic networks.

\end{abstract}

\maketitle

\section*{Introduction}
Cooperative photon emission can arise between quantum emitters because of photon-mediated interaction via a shared electromagnetic mode. This can occur with indistinguishable atoms, or artificial atoms, which emit identical photon wavepackets and cannot be spatially distinguished.
Atomic indistinguishability leads to entangled multi-particle states referred to as Dicke states~\cite{dicke_coherence_1954}.
In ensembles of densely packed atomic or solid-state emitters, Dicke states can yield sub- and superradiant emission with modified temporal, spectral, and directional properties~\cite{Gross_superadiance_1982,bradac_room-temperature_2017,rui_subradiant_2020,Angerer2018,Scheibner2007}. At the few emitter level, cooperative emission has been observed with emitters positioned closely (interatomic separation $\Delta < \lambda$, the photon wavelength) in free space~\cite{DeVoe1996,Eschner2001} or coupled to one-dimensional waveguides~\cite{vanLoo2013,sipahigil_integrated_2016,evans_photon-mediated_2018,kim_super-radiant_2018,grim_scalable_2019,prasad_correlating_2020}. Dicke states offer intriguing potential to engineer quantum states for applications in quantum information processing~\cite{Jen2016,Facchinetti_storing_2016,percel_topological_2017,asenjo_exponential_2017,shlesinger_time-frequency_2021} and precision metrology~\cite{krischek_useful_2011,pezze_quantum_2018}.

Atomic correlations can also occur from quantum interference between distant quantum emitters. This is commonly achieved by using a beam-splitter to erase the which-path information from two indistinguishable emitters, providing a route to realize scalable quantum networks~\cite{Cabrillo1999,Simon2003,Barrett2005,Moehring2007,bernien_heralded_2013,delteil_generation_2016}.
Similarly, interference in the far-field of spatially separated sources of indistinguishable single photons also gives rise to entanglement and Dicke states~\cite{Mandel1983,Wiegner2011,Wiegner2015}. Compared to the spontaneous emission from a single atom or a group of distinguishable atoms, a signature of the atomic entanglement that underlies cooperative emission is a change in the second-order intensity correlations; for instance, ``bunching'', rather than ``anti-bunching'', arises at zero delay in a Hanbury Brown–Twiss (HBT) interferometer~\cite{dolde_room-temperature_2013,sipahigil_integrated_2016,kim_super-radiant_2018,machielse_quantum_2019,grim_scalable_2019,wolf_light_2020,prasad_correlating_2020}. While cooperative emission, and in particular sub- and superradiance, have been extensively explored in both theory and experiment, coherent control of the correlations and the effects of indistinguishability and dephasing have yet to be investigated. Furthermore, higher-order intensity-intensity correlations that characterize the coherence~\cite{rezai_coherence_2018} of cooperative emission remain unexplored.

\begin{figure*}
  \includegraphics[width=1\textwidth]{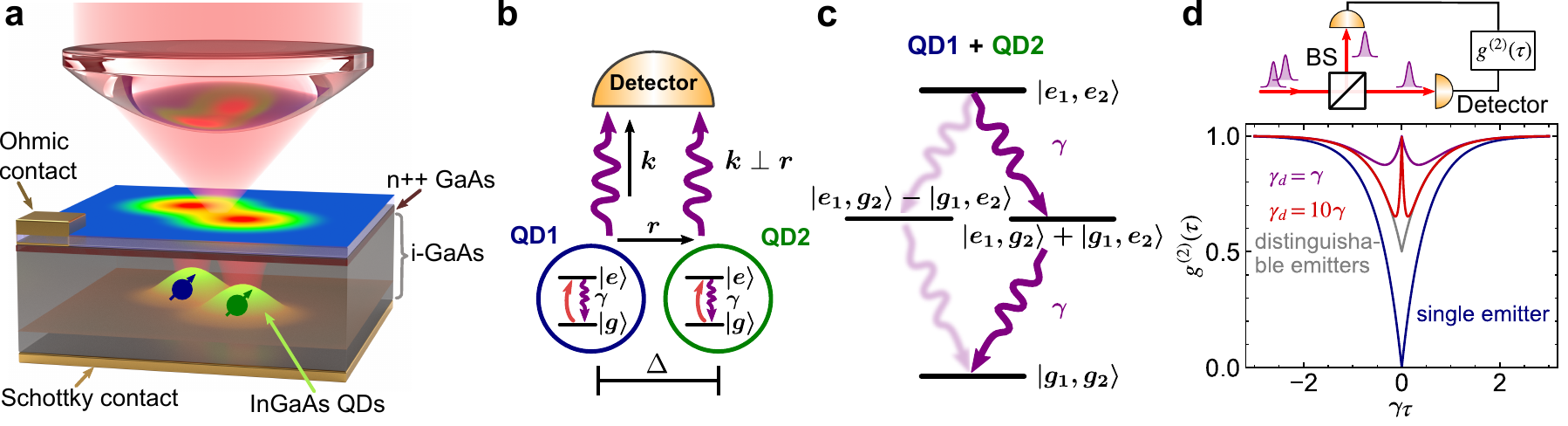}
  \caption{\textbf{Dicke states and cooperative emission from two indistinguishable quantum emitters.}
    \textbf{a} A schematic of the sample and spectroscopy setup, showing common-mode optical excitation and collection from two negatively charged InGaAs QDs embedded in an intrinsic (i-) GaAs region of a gate-tunable Schottky diode structure with an Ohmic contact to an n-doped (n++) GaAs layer.
    \textbf{b}
    Detection of photons with wave vector ($\bf{k}$) emitted from two indistinguishable QDs, spatially separated along ($\bf{r}$) by $\Delta$ and equally coupled to the same driving field, projects the system into the symmetric state ($\ket{e_1,g_2}+\ket{g_1,e_2}$). The spontaneous emission rate of a single QD is $\gamma$.
    \textbf{c} Schematic of the two-atom Dicke ladder, showing the bright transition from the doubly excited state ($\ket{e_1,e_2}$) to the symmetric state and subsequently to ground state ($\ket{g_1,g_2}$).
    Transitions via the anti-symmetric state ($\ket{e_1,g_2}-\ket{g_1,e_2}$, blurred arrows) are not directly monitored.
    \textbf{d} Second-order correlation function ($g^{(2)}(\tau)$) measured using the HBT setup (top) indicates the emergence of the ``anti-dip" around the zero delay due to cooperative emission from two indistinguishable emitters, each with a pure dephasing rate of $\gamma_d$.
    BS: 50:50 Beam splitter.
  }
  \label{fig:1}
\end{figure*}

Here, we report on the cooperative emission from two proximate semiconductor quantum dots (QDs) that can be tuned into resonance electrically. Collecting the emission using a diffraction-limited focus at $\Delta/2$ leads to the erasure of spatial and spectral distinguishability of the photons, creating emitter entanglement between the two dots, although the separation between them exceeds the wavelength of the emitted photons, $\Delta > \lambda$.
The ability to tune the emitter via an applied bias allows us to contrast the photon statistics of distinguishable to that of indistinguishable emitters.
Following the detection of a first photon, the detection probability of the subsequent photon is halved in the former case, because only one emitter remains excited and can contribute to photon emission.
By contrast, for cooperative emission from indistinguishable emitters, where both emitters are involved in both photon emission processes, the emission rates for the first and the second photon are identical.
As emission with a constant rate defines Poissonian statistics, the second-order correlation function then resembles that of a weak coherent state from an attenuated laser~\cite{rarity_non_2005}.

We probe the different statistics of distinguishable versus indistinguishable emission via second-order intensity correlations and with a Hong-Ou-Mandel (HOM) type interferometer~\cite{hong_measurement_1987}.
For both measurement configurations, we observe increased (zero-delay) correlations and Poisson-like photon statistics under continuous wave (CW) and pulsed driving.
Compared to strict resonance fluorescence, we observe a reduced level of correlation using nonresonant excitation (into the phonon sideband, blue-detuned from the zero-phonon line) due to increased emitter dephasing and time jitter of the exciton population.
Time-resolved resonance fluorescence of the independent (spectrally detuned) QDs and the correlated indistinguishable QD system reveals identical lifetimes, i.e.~we observe no reduction in the emission lifetime, as would be expected for superradiance~\cite{Gross_superadiance_1982}.
This indicates that the zero-delay peak in the HBT experiment is not a sufficient witness of superradiant emission with collective rate enhancements, but is a sensitive probe of cooperative emission.
Our work establishes new techniques from the quantum optics toolbox to control and characterize collective light-matter interactions.

Fig.~\ref{fig:1}(a) shows the schematic of our experiment: A confocal microscope with a diffraction-limited focus is used for both optical excitation and photon collection from two QDs.
Not shown in the schematic is a hemisphere solid immersion lens (SIL), with index of refraction $n = 2.0$, which is placed on top of the sample to increase the collection efficiency from the QDs~\cite{ma_efficient_2014}.
Crucially, the focal spot is optimized to ensure equal excitation and collection from both emitters, erasing the spatial distinguishability of each emitter. This is illustrated graphically in Fig.~\ref{fig:1}(b), in which two identical QDs, each with spontaneous emission rate $\gamma$, emit photons with wave vector $\mathbf{k}$ orthogonal to the separation between the emitters $\mathbf{r}$.
The indistinguishability of detected photons ensures that,
upon measurement of a single photon from a two-emitter system initially in the doubly-excited state $|e_1,e_2\rangle$, the wave function collapses to the maximally entangled Dicke state
$(1/\sqrt{2}\left(\ket{e_1,g_2}+\ket{g_1,e_2}\right))$
(cf. Fig.~\ref{fig:1}(c) and Materials and Methods).
The entanglement enables both emitters to cooperatively take part in the second photon emission process, which enhances photon coincidences compared to the emission from uncorrelated emitters. Here, the entanglement is induced by the measurement process. In contrast, entanglement between emitters in the superradiant regime arises from free radiative decay alone~\cite{dicke_coherence_1954, Gross_superadiance_1982}.
Fig.~\ref{fig:1}(d) shows examples of the second-order correlation function $g^{(2)}(\tau)$ near zero time delay (cf. section~S1) for CW driving of: (i) a single quantum emitter that exhibits perfect anti-bunching $g^{(2)}(0)=0$ (blue curve); (ii) two distinguishable quantum emitters that exhibit $g^{(2)}(0)=0.5$ (gray curve); and (iii) two indistinguishable quantum emitters with different dephasing rates $\gamma_d=\gamma$ and $10\,\gamma$ (purple and red curves, respectively).
An instructive example for which an analytic solution is available is the special case of incoherent pumping with equal pump and decay rates $\gamma_p=\gamma$, where one obtains a delay-time dependence of cooperative emission of the form
\begin{equation}
  g^{(2)}(\tau)=1- \big(e^{-2\gamma\tau} - e^{-(2\gamma+\gamma_d) \tau}\big)/2.
  \label{eqn:anti_dip_eqn}
\end{equation}
This yields the ``anti-dip" at $\tau=0$ with a width determined by the dephasing $\gamma_d$. As the main effect of dephasing here is to reduce the coherence between states with a single excitation in either emitter and therefore the correlations, the strong dependence of the $g^{(2)}(\tau)$ signal on the dephasing further highlights the importance of inter-emitter entanglement for the enhancement of photon coincidences.

\section*{Results}

\subparagraph{Tuning two near-degenerate QDs into resonance}

\begin{figure*}
  \includegraphics[width=1\textwidth]{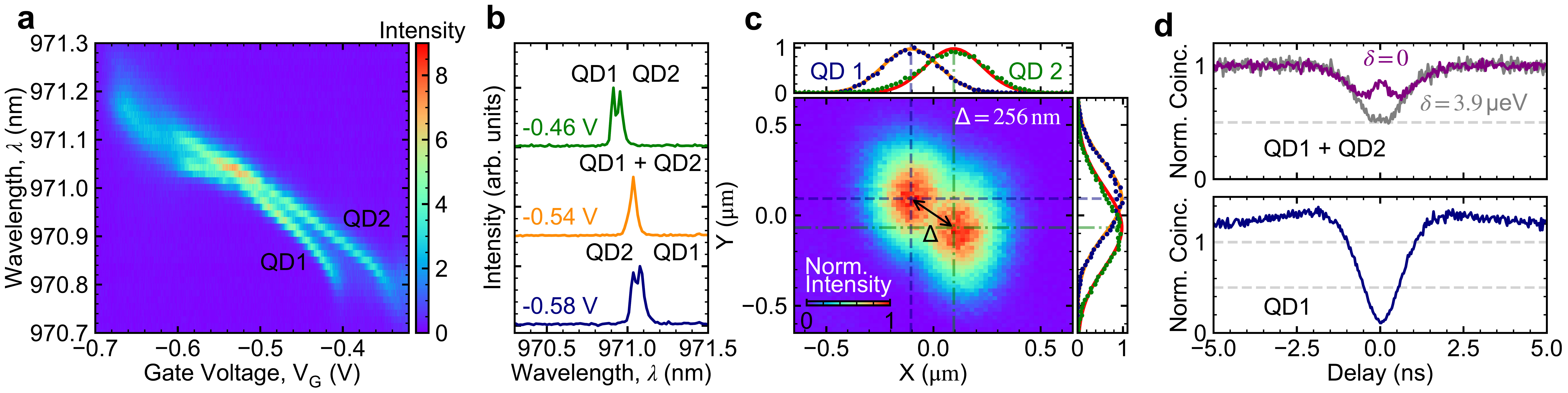}
  \caption{\textbf{Tuning two near-degenerate QDs into resonance.}
    \textbf{a} PL spectra (nonresonant excitation) of negatively charged exciton transitions from QD1 and QD2 as a function of applied gate voltage ($\mathrm{V_G}$).
    \textbf{b} Individual PL spectra at $\mathrm{V_G}=-0.46$, $-0.54$ and $-0.58\,\mathrm{V}$. The QDs are tuned into degeneracy at $\mathrm{V_G}=-0.54\,\mathrm{V}$.
    \textbf{c} Spatial profile of both QDs at $\mathrm{V_G}=-0.46\,\mathrm{V}$ accounting for the $\times 2$ magnification of the SIL.
    The Gaussian fits (solid lines) to the experimental data (circles) give the separation of two QDs of $\Delta=256\,\mathrm{nm}$.
    \textbf{d} Top: Second-order intensity correlation for emissions from both QDs (QD1+QD2) reveals an anti-dip around zero time delay for the case at zero detuning ($\delta=0$) and a dip showing a $g^{(2)}(0)\approx 0.5$ for nonzero detuning ($\delta=3.9\,\mu\mathrm{eV}$).
    Bottom: Second order intensity correlation ($g^{(2)}$) for emission from QD1 reveals dip at the zero delay, showing a $g^{(2)}(0)\approx 0.06$, signifying single-photon emission.
  }
  \label{fig:2}
\end{figure*}

The sample consists of self-assembled InGaAs QDs embedded in a Schottky diode to control the QD charge state via Coulomb blockade~\cite{warburton_optical_2000} and allow a small range of energy tuning via the DC Stark effect~\cite{warburton_giant_2002}. Details on the device design are found in Materials and Methods. The self-assembly process leads to random spatial positions and an inhomogeneous distribution of QD energies and spontaneous emission lifetimes ($\gamma$). We therefore search the sample for the unlikely situation in which two QDs: (i) are close enough to each other to optically couple to the same focal point, (ii) have very similar transition energies and $\gamma$, and (iii) have different permanent dipole moments such that the QDs can be tuned into resonance with a vertical electric field. We choose to work with negatively-charged exciton ($X^{1-}$) transitions, which unlike neutral excitons lack fine-structure splitting. Fig.~\ref{fig:2}(a) shows the photoluminescence (PL) (using nonresonant excitation) for our chosen QD pair ($X^{1-}$ transitions) as a function of gate bias. The three line cuts at different applied biases in Fig.~\ref{fig:2}(b) demonstrate the ability to tune the two QDs to the same emission wavelength using only the applied gate voltage. The resonance is found at $\mathrm{V_G}=-0.540~\mathrm{V}$. Here, $\lambda \approx 971$ nm in free space while inside the GaAs (with index of refraction $n \approx 3.67$), $\lambda_{\rm GaAs} \approx 265$ nm. Fixing $\mathrm{V_G}=-0.460~\mathrm{V}$ (where both QDs are spectrally distinguishable, detuned by $\delta \approx 70\,\mu \mathrm{eV}$), we spatially map the precise locations of the two QDs by scanning the sample position while recording the emission spectra. Assigning the peak at lower (higher) wavelength to be QD1 (QD2), we obtain the spatial profile of each QD, as shown in Fig.~\ref{fig:2}(c).
Gaussian fits to the intensity versus scanner position gives the spatial separation of the two QDs: $\Delta = 256.1\,(1)\,\mathrm{nm}$.
We note that for the two QDs in GaAs $\Delta \approx \lambda_{\rm GaAs}$, beyond the expected range for substantial superradiant emission enhancement~\cite{Gross_superadiance_1982}.
Here, we define the origin (X,Y)=(0,0) as the optimal spatial position to ensure equal collection from both QDs. For all subsequent experiments, we perform all measurements at this position and use the Stark shifts to render the emitters degenerate or detuned.

To confirm the indistinguishability and cooperative emission of the two QDs when degenerate, we obtain resonance fluorescence under CW driving and measure the second-order correlation function $g^{(2)}(\tau)$ with the HBT interferometer.
Fig.~\ref{fig:2}(d) shows $g^{(2)}(\tau)$ for three scenarios at an excitation power ($\mathrm{P}$) of $\mathrm{P/P_{sat}}\approx 0.1$, where $\mathrm{P_{sat}}$ is the excitation power at saturation~\cite{mollow_power_1969}. When the QDs are detuned and only QD1 is addressed resonantly, $g^{(2)}(0)\rightarrow 0$, indicating single photon emission. To equally drive the QDs when they are detuned ($\delta \approx 3.9\,\mu\mathrm{eV}$), we excite both emitters slightly off resonance with a laser detuned by $\pm \delta/2$, and obtain $g^{(2)}(0)\rightarrow 0.5$, as expected for two distinguishable emitters emitting uncorrelated single photons. Last, $g^{(2)}(0)$ for resonance fluorescence from the two degenerate QDs ($\delta = 0$) reveals the emergence of a zero-delay anti-dip, in agreement with previous reports for superradiant emission ~\cite{sipahigil_integrated_2016,kim_super-radiant_2018,grim_scalable_2019}.
This signature confirms cooperative emission from indistinguishable QDs.
We note that the slight bunching in coincidences away from the zero-delay for the single emitter case is due to spectral fluctuations. This is not observable for the two indistinguishable emitters case as their spectral fluctuations are not fully correlated.
We fit the experimental data to an analytical equation (c.f. Eq.~\ref{eqn:anti_dip_eqn}), convolved with the Gaussian instrument response function [full width at half maximum (FWHM) =0.240~ns] and obtain a decay time of $(2\gamma)^{-1}=0.880\,(22)\,\mathrm{ns}$, dephasing time of $\gamma_d^{-1}=0.199\,(10)\,\mathrm{ns}$ and hence a coherence time of $(2\gamma+\gamma_d)^{-1}=0.162\,(9)\,\mathrm{ns}$, indicating the 1/e width of the anti-dip.
While we observe $g^{(2)}(0)\approx 0.87$ from the data/fit, upon deconvolution we obtain $g^{(2)}(0)\approx 1$, as expected from collective emission from two indistinguishable emitters.

Last, as we increase the driving strength, we observe Rabi oscillations in the $g^{(2)}(\tau)$ coincidence histogram along with the zero-delay anti-dip. These results, depicted in section~S3, show excellent agreement with simulated data produced by our model in section~S1.

\subparagraph{Coherent control of cooperative emission}
\begin{figure*}
  \includegraphics[width=0.68\textwidth]{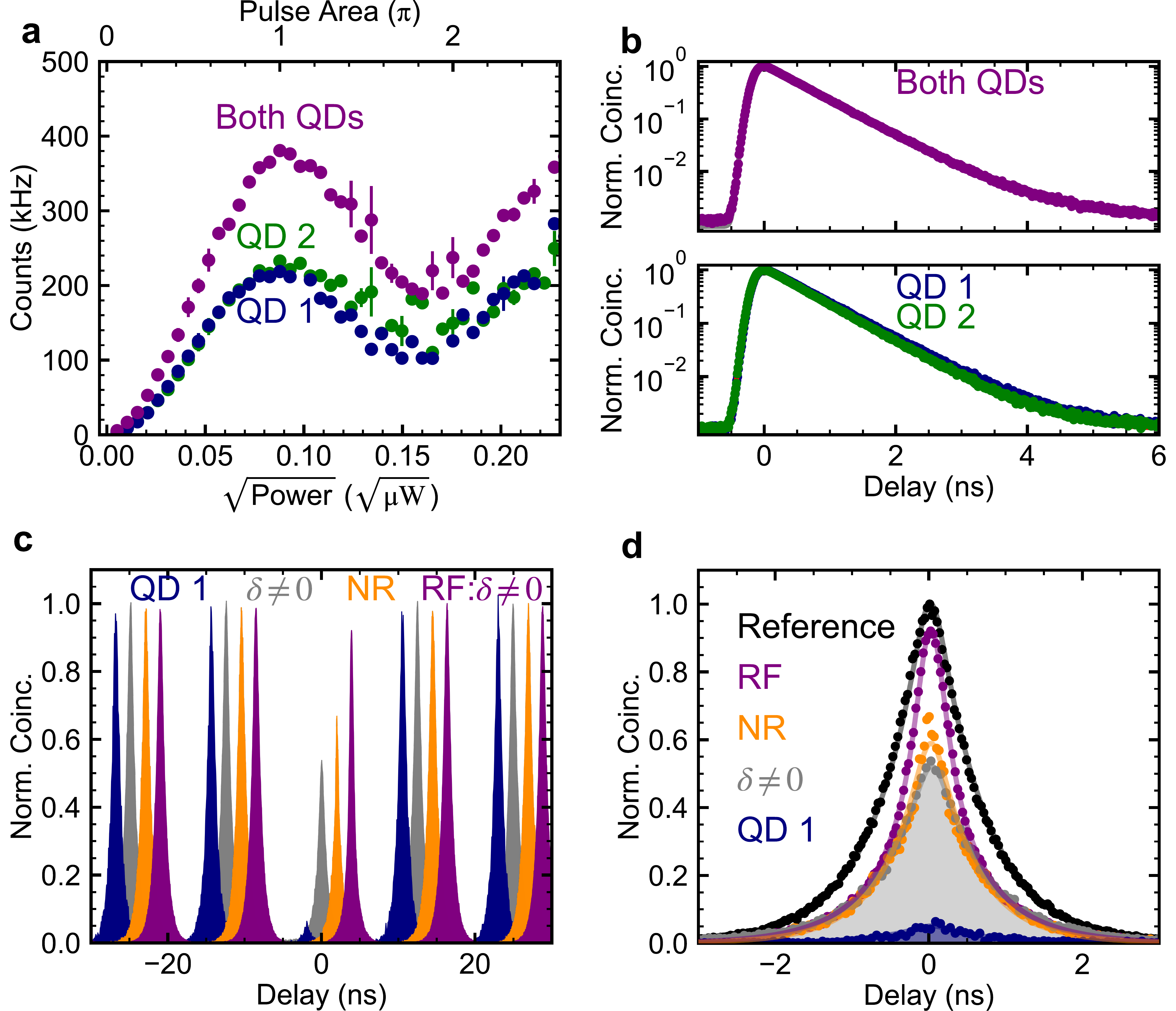}
  \caption{
    \textbf{Coherent control of two QDs.}
    \textbf{a} Emission intensity as a function of excitation power and pulse area from QD1 (blue), QD2 (green) and both QDs when degenerate (purple) under pulsed resonant fluorescence, RF.
    \textbf{b}
    Time-resolved emission profiles of the emission from the degenerate QDs (top) and individual emitters (bottom, QD1 and QD2) exhibit similar lifetimes of $T_1=0.643\,(1)$ and $0.668\,(1),~0.635\,(1)\,\mathrm{ns}$, respectively. The pulse area is $\pi/2$ for each measurement.
    \textbf{c} Second-order intensity correlation of the emission under resonant excitation [QD1 (blue), both QDs at non-zero detuning ($\delta\neq 0$, gray) and zero detuning ($\delta=0$, RF, purple)] and nonresonant excitation for the degenerate QDs (NR, orange).
    \textbf{d} Zoom-in of the zero-delay peaks in \textbf{c}.
    The data for ``Reference" are the side peak at 12.44~ns for the RF case.
    Measurements for RF, $\delta\neq0$ and QD1 are taken at the pulse area of $\pi/2$, while measurement for NR is taken at the saturation power.
  }
  \label{fig3}
\end{figure*}
To manipulate the Dicke ladder populations and probe the transient emission behavior, we perform resonance fluorescence and time-resolved measurements using a pulsed laser. Fig.~\ref{fig3}(a) shows coherent Rabi oscillations in the detected emission count rate as a function of pulse area for each emitter (when detuned) and for both QDs when degenerate. Each QD exhibits the same excitation power to Rabi frequency conversion and similar count rates, while the degenerate case yields the identical power to Rabi frequency conversion and the measured count rates are a sum of the individual QD count rates. Fixing the excitation power corresponding to a pulse area of $\pi/2$, we measure the temporal emission profile for each case with a time-resolved resonance fluorescence measurement, as illustrated in Fig.~\ref{fig3}(b). Here we observe nearly identical temporal profiles and lifetimes in each case: $T_1=0.668\,(1), 0.635\,(1)$, and $0.643\,(1)\,\mathrm{ns}$ for emission from QD1, QD2, and both QDs when degenerate, respectively. These results confirm that the temporal properties of the cooperative emission process are unchanged from the independent QDs; superradiance with a modified temporal profile is not observed, as expected for QDs with $\Delta \approx \lambda_{\rm GaAs}$.

Fig.~\ref{fig3}(c) presents $g^{(2)}(\tau)$ for resonance fluorescence at a pulse area of $\pi/2$ for four representative cases, with each dataset offset by a few nanoseconds on the x axis for clarity. A zoom-in around zero delay (without x-axis offset) is shown in Fig.~\ref{fig3}(d). Overall, the pulsed resonance fluorescence behaviour is analogous to that obtained under CW driving.
At zero delay, we observe $g^{(2)}(0) \rightarrow 0$, $g^{(2)}(0)\approx 0.5$, and $g^{(2)}(0)\rightarrow 1$ for an individual emitter (QD1), distinguishable emitters ($\delta\neq0$), and indistinguishable QDs (resonance fluorescence, RF), respectively.
For the individual-emitter case (QD1), the two QDs are detuned by $\sim 100\,\mu\mathrm{eV}$.
Integrating the coincidences within a 10~ns window around the zero delay, we obtain $g^{(2)}_{\Delta \tau =10\,\mathrm{ns}}(0)=0.051\,(1)$, $0.510\,(2)$ and $0.672\,(1)$ for each respective case. Restricting the integration window to 0.3~ns, while the $g^{(2)}_{\Delta \tau=0.3\,\mathrm{ns}}(0)$ values for emission from QD1 and distinguishable QDs remain the same, we instead obtain a much higher $g^{(2)}_{\Delta \tau=0.3\,\mathrm{ns}}(0)= 0.904\,(5)$ for two indistinguishable QDs, approaching the theoretical maximum of $1$. Comparison of these two integration windows of $g^{(2)}_{\Delta \tau}(0)$ for the degenerate QDs suggests that the cooperative emission is sensitive to the degree of indistinguishability or dephasing between the two emitters.
This effect is easily visualized by comparing the $g^{(2)}(0)$ peak for the indistinguishable QD (RF) to a reference peak (e.g. at $\tau$ = 12.44 ns, black plot), as shown in Fig.~\ref{fig3}(d).
Here, we observe narrowing of the zero-delay peak compared to the side peaks at $12.44\,\mathrm{ns}$.
We then fit the experimental data (zero-delay peak) with Eq.~\ref{eqn:center_peak_eqn} (convolved with the Gaussian instrument response function with FWHM=0.240~ns) to extract the dephasing parameter, $\gamma_d$.
\begin{equation}
  g^{(2)}_\mathrm{Pulsed}(\tau)=\big(e^{-\gamma \tau} + e^{-(\gamma +\gamma_d) \tau}\big)/2.
  \label{eqn:center_peak_eqn}
\end{equation}
Using the measured lifetime, $\gamma^{-1} = 0.643\,(1)\,\mathrm{ns}$,
we extract a dephasing time of $\gamma_d^{-1}  = 0.280\,(7)\,\mathrm{ns}$ and hence a coherence time of $(\gamma +\gamma_d)^{-1}=0.195\,(3)\,\mathrm{ns}$ from the fit.
By contrast, in the absence of dephasing ($\gamma_d=0$), the width of the zero-delay peak (given by Eq.~\ref{eqn:center_peak_eqn}) is the same as that of the side-peaks (equal to the radiative lifetime $\gamma^{-1}$), resulting in $g^{(2)}_{\Delta \tau}(0)=1$ for any integration time $\Delta_\tau$. This supports the interpretation that the degradation in $g^{(2)}_{\Delta \tau}(0)$ at larger $\Delta_\tau$ is related to emitter dephasing.

To experimentally demonstrate that the $g^{(2)}_{\Delta \tau}(0)$ value for cooperative emission is sensitive to dephasing, we excite the degenerate QDs using pulsed nonresonant, phonon-assisted excitation [NR, orange plot in Fig.~\ref{fig3}(c)] and observe reduced ``bunching" compared to strict resonance fluorescence: $g^{(2)}_{\Delta \tau = 0.3\,\mathrm{ns}}(0)= 0.624\,(7)$ and $g^{(2)}_{\Delta \tau = 10\,\mathrm{ns}}(0)= 0.556\,(4)$ within 0.3 and 10~ns integration windows, respectively.
In additional, fitting the data to Eq.~\ref{eqn:center_peak_eqn} results in a coherence time of $0.043\,(2)\,\mathrm{ns}$, significantly shorter than that observed with coherent driving.
We ascribe this reduction to reduced indistinguishability between the two QDs when using incoherent excitation, likely due to decoherence from the phonon-assisted transitions as well as increased charge noise and time jitter. Additional detailed spectroscopy results (power dependence and detuning dependence) under pulsed coherent and incoherent excitation are included in section~S4 and S5, respectively. In the case of coherent driving of two indistinguishable QDs, we observe an oscillating $g^{(2)}_{\Delta \tau}(0)$ value as we vary the excitation power of the resonant pulse, demonstrating the ability to coherently manipulate the populations on the Dicke ladder.

Last, we note that the bunching around zero delay demonstrates entanglement between the two QDs.
In the absence of detector jitter, the limit $g^{(2)}(\tau \to 0)$ indicates the degree of the instantaneous entanglement immediately after the first photon detection event as it is related to the occupation of the Dicke state (cf. Materials and Methods or section~S2).
However, in a realistic experimental setting, this limit is largely determined by finite detection jitter [with $g^{(2)}(0)=1$ for perfect detectors] and thus is not a reliable metric.
Therefore, we consider the integral of the entire zero-delay peak, relative to the adjacent uncorrelated side-peaks, as a more representative indicator of the degree of entanglement.
This quantity $g^{(2)}_{\Delta \tau}$ has the further advantage of being resilient to details of emitter dephasing that determine $g^{(2)}(\tau \to 0)$.
The degradation in $g^{(2)}_{\Delta \tau}$ due to the increase in the dephasing under nonresonant excitation (at saturation) results in a lower value of $g^{(2)}_{\Delta \tau=10\,\mathrm{ns}}\approx 0.56$ compared to that under resonant excitation $g^{(2)}_{\Delta \tau=10\,\mathrm{ns}}\approx 0.67$ (at a pulse area of $\pi/2$).
Correcting for both the laser leakage and the detection jitter, we obtain the associated entanglement fidelities (to the maximally entangled symmetric Dicke state), upon a detection of a single photon for both cases to be $\mathcal{F}\approx 0.56$ and $\mathcal{F}\approx 0.60$, respectively.
Please refer to section~S6 for a detailed discussion.

\subparagraph{Higher-order intensity correlations of cooperative emission}
\begin{figure*}
  \centering
  \includegraphics[width=\textwidth]{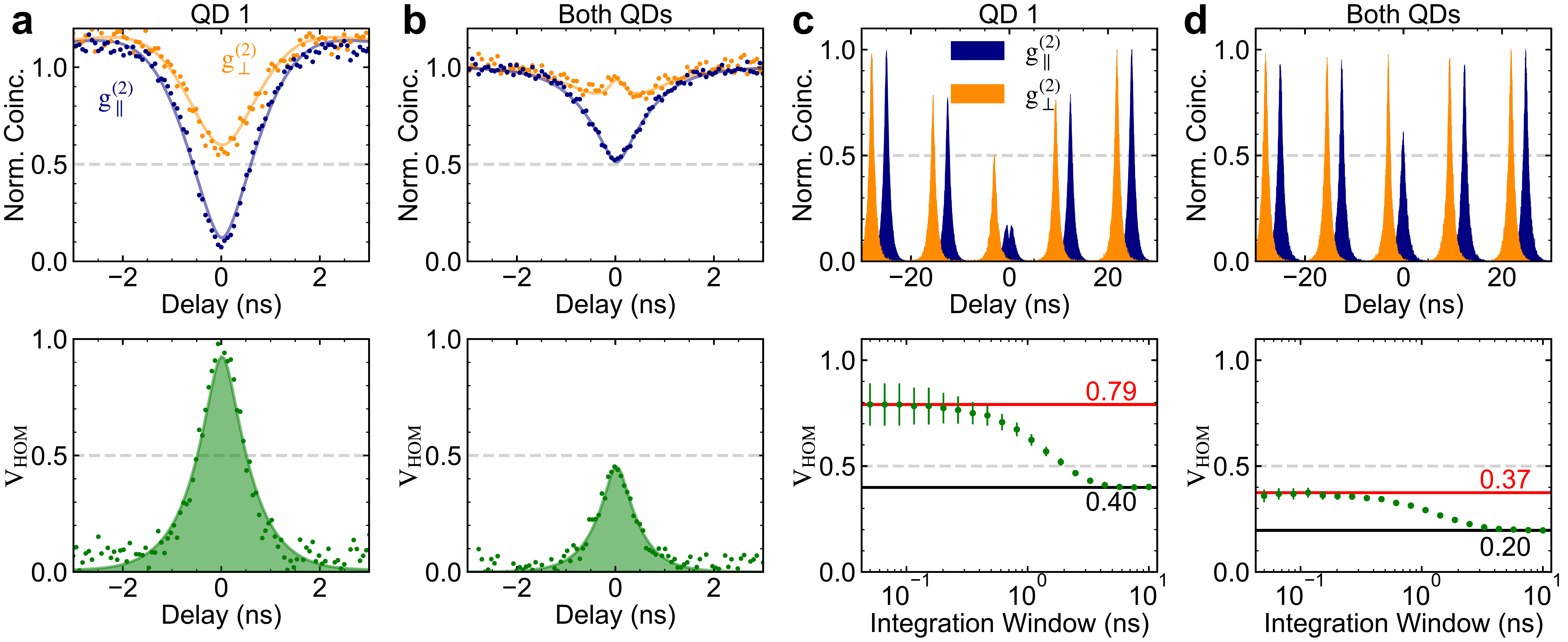}
  \caption{
  \textbf{HOM-type interferometry of cooperative emission.}
  \textbf{a, b} Normalized photon coincidences for the case where input photons from only QD1 (\textbf{a}) and from both QDs (\textbf{b}) are prepared in parallel ($g^{(2)}_\parallel$, blue) and perpendicular ($g^{(2)}_\perp$, orange) polarization, under CW resonant excitation.
  \textbf{c, d} Normalized photon coincidences ($g^{(2)}_\parallel$, $g^{(2)}_\perp$) for input photons from QD 1 \textbf{c} and both QDs \textbf{d}, under pulsed resonant excitation. The measurement in (\textbf{a}, \textbf{b}) and (\textbf{c}, \textbf{d}) is done at $\mathrm{P/P_{sat}}\approx 0.1$ and pulse area of $\pi/2$, respectively.
  Bottom: The corresponding interference visibility $\mathrm{V_{HOM}}$, calculated from the expression $1-g^{(2)}_\parallel/g^{(2)}_\perp$ for each case.
  }
  \label{fig4}
\end{figure*}

Next, we probe the coherence of the cooperative emission by interfering subsequently emitted photons in a conventional HOM-type interferometer setup. The measurement setup consists of an unbalanced Mach-Zehnder interferometer with a delay to match the temporal separation of the excitation pulses, i.e. $\Delta T = 12.44\,\mathrm{ns}$.
To quantify the interference visibility $\mathrm {V_{HOM}}(\tau)$, we define
\begin{equation}
  \mathrm {V_{HOM}}(\tau)=1-g^{(2)}_\parallel(\tau)/g^{(2)}_\perp(\tau),
  \label{eqn:1}
\end{equation}
where in the case of single-photon input, $\mathrm {V_{HOM}}(0)$ is the ratio of coincidences at zero delay when photons in the two paths of the interferometer are rendered indistinguishable ($g^{(2)}_\parallel(0)$) and distinguishable ($g^{(2)}_\perp(0)$) in polarization.
This polarization control of the input photons is achieved by rotating the half-wave-plate in one arm of the interferometer.
While $\mathrm {V_{HOM}}(\tau)$ gives the single-photon indistinguishability for a single-photon input, it provides a measure of the degree of coherence~\cite{rezai_coherence_2018} of the photons from both emitters.

Fig.~\ref{fig4}(a) and \ref{fig4}(b) show the results of the experiment for the emission from only QD1 and from both QDs when degenerate, respectively, under CW resonant driving. The temporal postselected indistinguishability for QD1 yields $\rm V_{HOM}(0)\rightarrow1$, as expected for an individual emitter. However, for cooperative emission from both QDs, we observe a maximum visibility of $\rm V_{HOM}(0)\approx0.5$.
This result signals a significant change in emitted light, from being anti-bunched (sub-Poissonian) for a single QD to having Poissonian-like statistics due to cooperative emission. Notably, the 1/e width of $\rm V_{HOM}$, which gives the coherence time of the emission, for the cooperative case ($\approx 1.15\,\mathrm{ns}$) is comparable to the single emitter case ($\approx 1.34\,\mathrm{ns}$).
Using the coherence time window (CTW)~\cite{proux_measuring_2015,baudin_correlation_2019,koong_fundamental_2019}, i.e. the integrated area of $\rm V_{HOM}$, as the figure of merit, we find that the CTW for the single-emitter case [$1.37\,(1)\,\mathrm{ns}$] is more than twice that of the cooperative emission [$0.53\,(1)\,\mathrm{ns}$], reflecting the difference for each case in both the $\rm V_{HOM}(0)$ values and 1/e widths.
Beyond the time-averaged picture, Fig.~\ref{fig4}(c) and \ref{fig4}(d) show the result of HOM0type interference for the emission from QD1 and both QDs under pulsed resonant driving, respectively. For QD1, we observe $\rm V_{HOM}=0.40\,(1)$ for the non-postselected (10~ns integration window) visibility and a maximum postselected (0.1~ns integration window) visibility of $\rm V_{HOM}=0.79\,(10)$.
These results indicate partial distinguishability between successively emitted photons from QD1. Similarly, compared to $\rm V_{HOM}=0.5$ obtained from the CW measurement and $\rm V_{HOM}=0.37\,(2)$ in the 0.1~ns integration window, degradation in the non-postselected visibility for the cooperative emission is observed: $\rm V_{HOM}=0.20\,(1)$ within a 10~ns integration window.

\section*{Discussion}
In summary, we coherently control and probe two QDs coherently coupled by a photon-mediated interaction enabled by the indistinguishability of their photon wavepackets and path erasure of their spatial positions.
We find that both $g^{(2)}(\tau)$ and lifetime measurements are necessary to fully describe the nature of the coherent coupling; $g^{(2)}(\tau)$ on its own is not a sufficient witness of superradiance.
Compared to incoherent driving, we demonstrate that coherent driving enables both increased emitter indistinguishability and control of the populations on the Dicke ladder. We find that after emission of the first photon, both emitters are in an entangled state, as signified by $g^{(2)}(0)\rightarrow 1$. Furthermore, we show that $g^{(2)}(\tau)$ is a sensitive probe of the dephasing of the emitters. For coherent driving, the dephasing likely originates from emitter coupling to the solid-state environment [phonon-induced dephasing~\cite{ramsay_phonon-induced_2010,koong_fundamental_2019,brash_light_2019} or charge and spin noise~\cite{kuhlmann_charge_2013}], which can be at least partially mitigated with improved device quality, choice of charge state, and Purcell enhancement.
Last, our measurements reveal Poissonian-like statistics of the cooperative emission, analogous to a weak coherent state from an attenuated laser~\cite{rarity_non_2005}.
This result demonstrates the coherence of the cooperative emission and evokes the analogy between cooperative emission, in which indistinguishable atomic dipoles are locked in-phase, and lasing~\cite{Gross_superadiance_1982}.
Exciting prospects would be to increase the number of cooperative emitters using scalable approaches and include spin control of the coherently coupled emitters. This would create more intermediate states on the Dicke ladder, allow better understanding of the collective behaviour of interacting many-body systems, and provide a potential route to tailor the cooperative behaviour and harness collective light-matter interaction effects for photon-mediated applications.

\section*{Acknowledgements}
\textbf{Funding:} This work was supported by the EPSRC (grant no. EP/L015110/1, EP/M013472/1, EP/P029892/1, and EP/T01377X/1), the ERC (grant no. 725920), and the EU Horizon 2020 research and innovation program under Grant Agreement No. 820423.
B. D. G. thanks the Royal Society for a Wolfson Merit Award and the Royal Academy of Engineering for a Chair in Emerging Technology,
T.S.S. acknowledges PNPD/CAPES for financial support.
The authors in K.I.S.T. acknowledge the program of quantum sensor core technology through IITP (MSIT grant no. 20190004340011001).
\textbf{Author contributions: } B.D.G. and E.M.G. conceived and supervised the project. S.I.P. and J.D.S. fabricated the samples. Z.X.K. and T.S.S. performed the experiments. M.C., E.S. and T.S.S. developed the theoretical model and performed numerical simulations.
Z.X.K., M.C., E.M.G., and B.D.G. cowrote the paper with input from all authors.
\textbf{Competing interests:} The authors declare that they have no competing interests.
\textbf{Data and materials availability:} All data needed to evaluate the conclusions in the paper are present in the paper and/or the Supplementary Materials.
Additional data related to this paper is available on \url{https://doi.org/10.17861/638f5d04-9042-4215-9a0a-fc4d09a9c036}.

% \clearpage
\newpage

\section*{Materials and methods}
\subparagraph{Sample structure}
The experiments are performed on self-assembled InGaAs QDs embedded in a GaAs Schottky diode for deterministic charge control via the applied gate voltage.
The gate voltage induces DC Stark shift and shifts the emission wavelength of the emitter, with a typical linear response of $1$ to $2\,\mu \mathrm{eV/mV}$.
A broadband planar cavity and glass solid immersion lens (not shown in Fig.~\ref{fig:1}(a)) are used to enhance the photon extraction efficiency~\cite{ma_efficient_2014}.
Specific details of the sample heterostructure are described in~\cite{dada_indistinguishable_2016}.

\subparagraph{Spectroscopy setup}
The sample is kept at a temperature of $4\,\mathrm{K}$ in a closed-cycle helium flow cryostat.
A polarization-based dark-field confocal microscope is used to excite and collect the resonance fluorescence from the QDs while suppressing the scattered laser light (typical extinction ratio $\approx 10^7$)~\cite{kuhlmann_dark_2013}.
The photons are sent to either a spectrometer (resolution of $40\,\mu\mathrm{eV}$) equipped with a liquid nitrogen-cooled charge-coupled device or a pair of superconducting nanowire single photon detectors with a nominal detection efficiency of $90\,\%$ at 950~nm and a time jitter of $\approx 100$~ps. For HOM measurements, we spectrally filter the phonon sideband using a grating filter with a FWHM of $120\,(1)\,\mu \mathrm{eV}$ and $45-50\,\%$ fiber-to-fiber transmission efficiency.
The CW experiments are done with a narrow band (sub-megahertz linewidth) diode laser while the pulsed driving is achieved with a mode-locked femtosecond laser, stretched to $\sim 40\,\mathrm{ps}$ pulse width, with an 80~MHz repetition rate.
Emitter lifetime and photon correlations (HBT and HOM) are made with a time-correlated single photon counting system.

\subparagraph{Coincidence measurements under cooperative emission}
Photon coincidences provide a useful tool to characterize cooperative emission. To derive the signatures of cooperative emission, we consider the light-matter interaction between two QDs described as two-level systems, where $|g_i\rangle$ and $|e_i\rangle$ are ground and excited states of the $i$-th QD.
Introducing the QD operators $\sigma_i^+=|e_i\rangle\langle g_i|$ and $\sigma_i^-=|g_i\rangle\langle e_i|$ as well as photon creation and annihilation operators $a^\dagger_\mathbf{k}$ and $a_\mathbf{k}$, the light-matter interacting Hamiltonian is
\begin{align}
  H_I= & \sum_\mathbf{k}\hbar g \Big[
    \big( e^{i\mathbf{k}\cdot\mathbf{r}/2} \sigma^-_1
    + e^{-i\mathbf{k}\cdot\mathbf{r}/2} \sigma^-_2 \big) a^\dagger_\mathbf{k}
    + h.c. \Big].
\end{align}
If the two QDs are fully distinguishable, then the emission of
photons can be described by nondegenerate perturbation theory.
The signal registered at the detectors is then simply a sum
of the individual detection events
\begin{align}
  g^{(2)}_\textrm{indep.}(t,\tau)=
  \frac{\sum\limits_{i,j=1}^2 \langle \sigma^+_i(t)\sigma^+_j(t+\tau)
    \sigma^-_j(t+\tau)\sigma^-_i(t)\rangle}
  {\sum\limits_{i,j=1}^2 \langle \sigma^+_i(t)\sigma^-_i(t)\rangle
    \langle \sigma^+_j(t+\tau)\sigma^-_j(t+\tau)\rangle}.
\end{align}
Following the emission of a photon, one QD will be in its ground
state and can thus no longer contribute to the immediate emission of a second photon, i.e.~$\sigma^-_{i}\sigma^-_i=0$. Hence, only emission processes from different emitters
$i\neq j$ contribute to the numerator, whereas all emission processes contribute to the denominator.
For initially uncorrelated and distinguishable QDs with exciton populations $n_1$ and $n_2$ at time $t$, the zero-delay coincidences are therefore
\begin{align}
  g^{(2)}_\textrm{indep.}(t,0)=
  \frac{2n_1n_2}{(n_1+n_2)^2}\le \frac 12,
\end{align}
where the limit $1/2$ is reached for $n_1=n_2$.

However, when the QDs are indistinguishable, both QDs feed into the same electromagnetic field modes and a signal measured at the detector can no longer resolve from which QD a detected photon originated.
To describe the measurement process, it is instructive to rewrite the light-matter interaction
\begin{align}
  H_I= &
  \sum_\mathbf{k}\hbar g\sqrt{2} \big(
  \sigma^-_\mathbf{k} a^\dagger_\mathbf{k} + \sigma^+_\mathbf{k}a_\mathbf{k}
  \big)
  \label{eq:HIrewrite}
\end{align}
in terms of operators
$\sigma^\pm_\mathbf{k}=\big(
  e^{i\mathbf{k}\cdot\mathbf{r}/2} \sigma^\pm_1
  + e^{-i\mathbf{k}\cdot\mathbf{r}/2} \sigma^\pm_2 \big)/\sqrt{2}$,
which describe the dipole operator responsible for driving the photon mode with wave vector $\mathbf{k}$.

In our setup, we predominantly detect photons emitted perpendicular to the plane containing the QDs, i.e. photons with wave vectors $\mathbf{k}\approx \mathbf{k}_0$, where $\mathbf{k}_0$ is a reference wave vector with $\mathbf{k}_0\perp \mathbf{r}$ and
modulus $|\mathbf{k}_0|$ matching the common transition frequency of the QDs.
In this situation, $e^{\pm i\mathbf{k}\cdot\mathbf{r}}\approx 1$ so that the photon modes picked up by the detector are driven by the dipole operators $\sigma_S^\pm=\big( \sigma^\pm_1 + \sigma^\pm_2 \big)/\sqrt{2}$, describing transitions through the symmetric Dicke state
$|\psi_S\rangle = \big(|e_1,g_2\rangle+|g_1,e_2\rangle\big)/\sqrt{2}$
[cf. Fig.~\ref{fig:1}(c)].
In particular, the second-order coherence takes the form
\begin{align}
  g^{(2)}_\textrm{coop.}(t,\tau)= & \frac{\langle \sigma^+_S(t)\sigma^+_S(t+\tau)
    \sigma^-_S(t+\tau)\sigma^-_S(t)\rangle}
  {\langle \sigma^+_S(t)\sigma^-_S(t)\rangle
    \langle \sigma^+_S(t+\tau)\sigma^-_S(t+\tau)\rangle} .
\end{align}

With occupation $n_{e_1,e_2}$ of the doubly excited state $|e_1,e_2\rangle$ and occupations $n_S$ of the symmetric state $|\psi_S\rangle$,
the zero-delay coincidences are
\begin{align}
  g^{(2)}_\textrm{coop.}(t,0)=\frac{n_{e_1,e_2}}{(n_{e_1,e_2}+n_S)^2},
\end{align}
which for initially uncorrelated states with equal exciton populations results in $g^{(2)}_\textrm{coop.}(t,0)=1$.

The decisive difference to the situation of independent emitters is that the emission of the first photon does not impede the emission of a second photon.
Instead, both emitters cooperatively contribute to both emission processes, leading
to zero-delay coincidences that exceed the limit of $\frac{1}{2}$ for independent emitters.

A minimal model for calculating the time dependence of $g^{(2)}(t,\tau)$ is given in section~S1.

\section*{Supplementary materials}
Supplementary material for this article is available at \url{https://science.org/doi/10.1126/sciadv.abm8171}.

\bibliography{reference}
\end{document}

\title{Supplementary material for coherence in cooperative photon emission from indistinguishable quantum emitters}

\author{Zhe Xian Koong}
\email[Correspondence: ]{zk49@hw.ac.uk}
\affiliation{
  SUPA, Institute of Photonics and Quantum Sciences, Heriot-Watt University, Edinburgh EH14 4AS, UK.
}
\author{Moritz Cygorek}
\author{Eleanor Scerri}
\affiliation{
  SUPA, Institute of Photonics and Quantum Sciences, Heriot-Watt University, Edinburgh EH14 4AS, UK.
}
\author{Ted S. Santana}
\affiliation{
Centro de Ci\^{e}ncias Naturais e Humanas, Universidade Federal do ABC, Santo Andr\'{e}, São Paulo 09210-580, Brazil.
% Departamento de F\'{i}sica, Universidade Federal de Sergipe, 49100-000, Brazil
}
\author{Suk In Park}
\affiliation{
  Center for Opto-Electronic Materials and Devices Research, Korea Institute of Science and Technology, Seoul 02792, Republic of Korea
}
\author{Jin Dong Song}
\affiliation{
  Center for Opto-Electronic Materials and Devices Research, Korea Institute of Science and Technology, Seoul 02792, Republic of Korea
}
\author{Erik M. Gauger}
\author{Brian D. Gerardot}
\email[Correspondence: ]{b.d.gerardot@hw.ac.uk}
\affiliation{
  SUPA, Institute of Photonics and Quantum Sciences, Heriot-Watt University, Edinburgh EH14 4AS, UK.
}
\email{zk49@hw.ac.uk}

\date{March 18, 2022}

\maketitle

\renewcommand\thesection{S\arabic{section}}
\renewcommand{\thefigure}{S\arabic{figure}}
\renewcommand{\theequation}{S\arabic{equation}}
\setcounter{equation}{0}
\setcounter{section}{0}
\setcounter{figure}{0}
\setcounter{page}{1}

\section{Minimal model for $g^{(2)}(\tau)$}\label{appendix:minimalmodel}
\begin{table*}[h]
  \bgroup
  \def\arraystretch{1.5}
  \begin{tabular}{|c||c|c|}
    \hline
    % {\bf Emission model:} &$\frac{\partial}{\partial t}\rho=$ & $G^{(2)}(\tau)=$ \\
    {\bf Emission model:} & $\partial\rho/\partial t=$ & $G^{(2)}(\tau)=$ \\
    \hline
    {\bf single}          &
    \begin{minipage}{7.4cm}
      \vspace{2mm}
      $\gamma \mathcal{L}\big[\sigma^-_1\big](\rho)
        +\gamma_p \mathcal{L}\big[\sigma^+_1\big](\rho)
        +\gamma_d \mathcal{L}\big[\sigma^+_1\sigma^-_1\big](\rho)$
      \vspace{2mm}
    \end{minipage}
                          &
    $\langle \sigma^+_1 \sigma^+_1(\tau)\sigma^-_1(\tau)\sigma^-_1\rangle$
    \\
    \hline
    {\bf independent}     &
    \begin{minipage}{7.4cm}
      \vspace{2mm}
      $\gamma \mathcal{L}\big[\sigma^-_1\big](\rho)
        +\gamma_p \mathcal{L}\big[\sigma^+_1\big](\rho)
        +\gamma_d \mathcal{L}\big[\sigma^+_1\sigma^-_1\big](\rho)$\\[1mm]$
        +\gamma \mathcal{L}\big[\sigma^-_2\big](\rho)
        +\gamma_p \mathcal{L}\big[\sigma^+_2\big](\rho)
        +\gamma_d \mathcal{L}\big[\sigma^+_2\sigma^-_2\big](\rho)$
      \vspace{0mm}
    \end{minipage}
                          &
    $\sum\limits_{i,j=1,2}\langle \sigma^+_i\sigma^+_j(\tau)\sigma^-_j(\tau)\sigma^-_i\rangle$
    \\
    \hline
    {\bf cooperative}     &
    \begin{minipage}{7.4cm}
      \vspace{2mm}
      $\gamma \mathcal{L}\big[\sigma^-_1\big](\rho)
        +\gamma_p \mathcal{L}\big[\sigma^+_1\big](\rho)
        +\gamma_d \mathcal{L}\big[\sigma^+_1\sigma^-_1\big](\rho)$\\[1mm]$
        +\gamma \mathcal{L}\big[\sigma^-_2\big](\rho)
        +\gamma_p \mathcal{L}\big[\sigma^+_2\big](\rho)
        +\gamma_d \mathcal{L}\big[\sigma^+_2\sigma^-_2\big](\rho)$
      \vspace{0mm}
    \end{minipage}
                          &
    $\langle \sigma^+_S\sigma^+_S(\tau)\sigma^-_S(\tau)\sigma^-_S\rangle$
    \\
    \hline
    {\bf superradiant}    &
    \begin{minipage}{7.4cm}
      \vspace{2mm}
      $\Gamma \mathcal{L}\big[\sigma^-_S\big](\rho)
        +\gamma_p \mathcal{L}\big[\sigma^+_1\big](\rho)
        +\gamma_d \mathcal{L}\big[\sigma^+_1\sigma^-_1\big](\rho)$\\[1mm]
      $+\gamma_p \mathcal{L}\big[\sigma^+_2\big](\rho)
        +\gamma_d \mathcal{L}\big[\sigma^+_2\sigma^-_2\big](\rho)
      $
      \vspace{2mm}
    \end{minipage}
                          &
    $\langle \sigma^+_S\sigma^+_S(\tau)\sigma^-_S(\tau)\sigma^-_S\rangle$
    \\
    \hline
  \end{tabular}
  \egroup
  \caption{\label{tab:eom}Equations of motion and expressions for the second
    order (intensity) correlation function $G^{(2)}$ for
    the cases of a single emitter, independent emitters, selectively measured
    emitters, and superradiantly coupled emitters.}
\end{table*}
Here, we formulate a minimal model to describe the time delay dependence
of $g^{(2)}(\tau)$ for single emitters, independent emitters, selectively
measured coherently coupled emitters, and superradiantly coupled emitters.
To this end, we need to calculate two-time correlation functions of the form
\begin{align}
  \begin{split}
    %G^{(2)}_{\mathbf{k}_1,\mathbf{k}_2}(t_1,\tau)&=
    &\langle a^\dagger_{\mathbf{k}_1}(t_1) a^\dagger_{\mathbf{k}_2}(t_1+\tau)
    a_{\mathbf{k}_2}(t_1+\tau) a_{\mathbf{k}_1}(t_1)\rangle\\
    &=\textrm{Tr}\Big[ a_{\mathbf{k}_2}(t_1+\tau) a_{\mathbf{k}_1}(t_1)\rho(0) a^\dagger_{\mathbf{k}_1}(t_1) a^\dagger_{\mathbf{k}_2}(t_1+\tau) \Big]\\
    &=\textrm{Tr}\Big[ a_{\mathbf{k}_2} U(\tau) \big(a_{\mathbf{k}_1}
      \rho(t_1) a^\dagger_{\mathbf{k}_1} \big)U^\dagger (\tau)a^\dagger_{\mathbf{k}_2} \Big].
  \end{split}
  \label{eq:G2calc}
\end{align}
This can be achieved by propagating the density matrix $\rho$ using the
time evolution operator $U(t)$ up to time $t_1$, applying the photon operators
$a^{(\dagger)}_{\mathbf{k}_1}$, and propagating the resulting non-normalized pseudo density matrix for a time $\tau$, before applying the second set of photon operators
$a^{(\dagger)}_{\mathbf{k}_2}$.
The photon operators $a_\mathbf{k}$ are replaced by system operators
$\sigma^-_i$ or $\sigma^-_S$ depending on the situation (cf. main text).

In our minimal model, we describe the time evolution before and in between the photon detection events using incoherent pump, decay, and dephasing rates.
Because the selective measurement or superradiant decay introduces coherence between excited states in the two emitters, we use master equations of Lindblad form to correctly capture the corresponding off-diagonal elements in the density matrix, replacing the time evolution operators $U(t)$ in Eq.~\eqref{eq:G2calc} by the corresponding propagator in Liouville space.

Using the Lindblad superoperator
\begin{align}
  \mathcal{L}[A](\rho)=A\rho A^\dagger
  -\tfrac 12 \big(A^\dagger A\rho + \rho A^\dagger A\big),
\end{align}
the time evolution of a single incoherently pumped emitter can be written as
\begin{align}
  \frac{\partial }{\partial t} \rho= &
  \gamma \mathcal{L}\big[\sigma^-_1\big](\rho)
  +\gamma_p \mathcal{L}\big[\sigma^+_1\big](\rho)
  +\gamma_d \mathcal{L}\big[\sigma^+_1\sigma^-_1\big](\rho),
\end{align}
where the first term describes radiative decay with rate $\gamma$,
the second term accounts for incoherent pumping with rate $\gamma_p$, and
the third term phenomenologically describes pure-dephasing with a
dephasing rate $\gamma_d$. The latter has the effect of reducing coherence
between the ground $|g_1\rangle$ and excited state $|e_1\rangle$ and originates
microscopically, e.g., from interactions with the phonon environment.

The equations of motion for two independent emitters are obtained
by adding to the equations of motion for a single emitter the same set of
equations but with operators $\sigma^\pm_1$ replaced by $\sigma^\pm_2$.
In the case of superradiance, we assume the pumping of both emitters to
be independent, but radiative decay involves transitions through the
symmetric state
\begin{align}
  \frac{\partial }{\partial t} \rho= &
  \Gamma \mathcal{L}\big[\sigma^-_S\big](\rho)
  +\gamma_p \mathcal{L}\big[\sigma^+_1\big](\rho)
  +\gamma_p \mathcal{L}\big[\sigma^+_2\big](\rho),
\end{align}
where for ideal (no pure dephasing) superradiance $\Gamma=2\gamma$.
The respective equations of motion for the different cases together with the corresponding expressions for the second order coherence are listed in Table~\ref{tab:eom}.
With these results, the numerical calculation of the second order correlation function is
straightforward. If not stated otherwise, we use parameters
$\gamma_p=\gamma$ and $\gamma_d=0$, where $\gamma=1$ defines the unit of time.

\begin{figure}
  \includegraphics[width=0.5\textwidth]{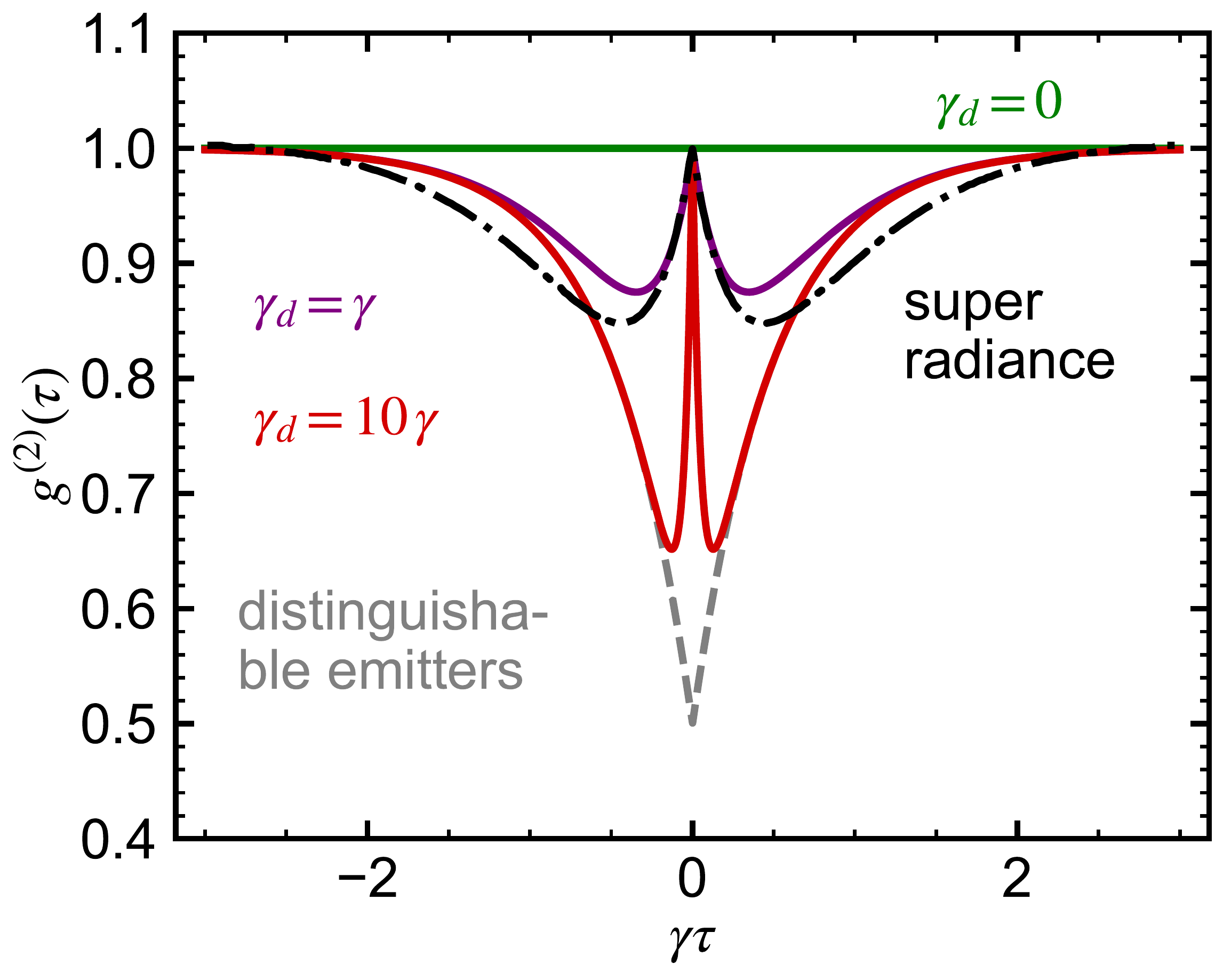}
  \caption{
    \textbf{Second-order correlation function ($g^{(2)}(\tau)$) for selective measurement and superradiant case.}
    $g^{(2)}(\tau)$ as a function of detection time delay, $\tau$ shows similar anti-dip feature around $\tau=0$ as a result of emission from two indistinguishable emitters, under the selective measurement ($\gamma_d\neq 0 $, solid lines) and superradiance (dashed-dotted line).
    In the absence of pure dephasing ($\gamma_d =0$, solid line), $g^{(2)}(\tau)=1$ at all $\tau$.
    For the case of two distinguishable emitters (dashed line), this feature disappears and the $g^{(2)}(0)=0.5$ instead.
  }
  \label{fig:supmat_theory1}
\end{figure}

\section{Relation to Superradiance}

To further clarify the nature of cooperative emission in absence of superradiance, we now describe the link to superradiance on the one hand and to independent emission on the other hand in terms of wave vector selectivity of coincidence measurements.

Consider two detectors that predominantly register photons at wave vectors $\mathbf{k}_0^{(1)}$ and $\mathbf{k}_0^{(2)}$, respectively. Then, a first photon detection event described by the operators $\sigma^\pm_{\mathbf{k}_0^{(1)}}$ from an initially doubly occupied two-emitter system leads to a collapse of the emitter state to $|\psi_{\mathbf{k}_0^{(1)}}\rangle$.
The conditional probability of detecting a second photon by detector 2 as described by operators $\sigma^\pm_{\mathbf{k}_0^{(2)}}$ is given by
\begin{align}
  G^{(2)}_{\mathbf{k}_0^{(1)},\mathbf{k}_0^{(2)}}(t,0)= &
  \langle \sigma^+_{\mathbf{k}_0^{(1)}}\sigma^+_{\mathbf{k}_0^{(2)}} \sigma^-_{\mathbf{k}_0^{(2)}}\sigma^-_{\mathbf{k}_0^{(1)}}\rangle
  =
  \langle \psi_{\mathbf{k}_0^{(1)}}|\sigma^+_{\mathbf{k}_0^{(2)}} \sigma^-_{\mathbf{k}_0^{(2)}}|\psi_{\mathbf{k}_0^{(1)}}\rangle
  =
  |\langle \psi_{\mathbf{k}_0^{(2)}}|\psi_{\mathbf{k}_0^{(1)}}\rangle|^2=\frac 12\Big[1+\cos\big( (\mathbf{k}_0^{(2)}-\mathbf{k}_0^{(1)})\cdot \mathbf{r}\big)\Big],
  \label{eq:suppl_Gkk}
\end{align}
which shows a strong dependence of photon coincidences on the relative emission directions
of the first and the second photon. Eq.~\eqref{eq:suppl_Gkk} can be interpreted as follows: The first photon measurement process prepares the system in a state whose oscillator strength for the emission of a second photon strongly depends on the emission direction $\mathbf{k}_0^{(2)}$, encoded in the relative phase $e^{i\mathbf{k}_0^{(1)}\cdot\mathbf{r}}$ between states $|e_1,g_2\rangle$ and $|g_1,e_2\rangle$ in the correlated intermediate state
$|\psi_{\mathbf{k}_0^{(1)}}\rangle$.

In the superradiant regime $\mathbf{k}_0^{(i)}\cdot\mathbf{r}\approx0$, so that the states involved in the transitions have identical phases
$|\psi_{\mathbf{k}_0^{(2)}}\rangle=|\psi_{\mathbf{k}_0^{(1)}}\rangle=|\psi_S\rangle$,
leading to a perfect overlap $|\langle \psi_{\mathbf{k}_0^{(2)}}|\psi_{\mathbf{k}_0^{(1)}}\rangle|^2=1$ and hence a maximal value of
$G^{(2)}_{\mathbf{k}_0^{(1)},\mathbf{k}_0^{(2)}}(t,0)=1$, irrespective of the emission directions defined by $\mathbf{k}_0^{(1)}$ and $\mathbf{k}_0^{(2)}$.

By contrast, if the distance $r$ is larger than the wavelength of the emitted light and all photons are collected, i.e. no particular wave vectors are selected, then the relative phases
$e^{i\mathbf{k}_0^{(i)}\cdot\mathbf{r}}$ are random, resulting in destructive interference of the cosine term in Eq.~\eqref{eq:suppl_Gkk}. This leads to exactly half as many photon coincidences as in the superradiant regime, analogous to what we obtain from simply considering independent emitters.

Finally, in our confocal microscope, we predominantly detect photons emitted
perpendicular to the sample plane containing the QDs, which implies that all detected photons have similar wave vectors $\mathbf{k}_0^{(2)}=\mathbf{k}_0^{(1)}$. Therefore, with
$\cos\big( (\mathbf{k}_0^{(2)}-\mathbf{k}_0^{(1)})\cdot \mathbf{r}\big)=1$, the same enhancement of $G^{(2)}(t,0)=1$ is found as in the superradiant regime. This remains true
even though the distance between the emitters may be significantly larger than the wavelength of the emitted light.
In both of the `cooperative' cases, i.e.~ superradiance as well as the present case of cooperative emission without superradiance, the enhancement of photon coincidences with respect to independent emitters can be attributed to the perfect matching between intermediate states involved in the emission processes of the first and the second photon.

\commentout{
A well-known effect of cooperative emission is superradiance~\cite{Gross_superadiance_1982}
which arises when multiple emitters are confined to a region smaller than the
wavelength of the emitted light $\mathbf{k}\cdot\mathbf{r}\to 0$.
Consequently, photons couple to the same dipole
$\sigma_S^\pm=\frac 1{\sqrt{2}}\big( \sigma^\pm_1 + \sigma^\pm_2 \big)$
irrespective of the emission direction, so that radiative decay always involves
transitions through the symmetric Dicke state
$|\psi_S\rangle =\frac 1{\sqrt{2}} \big(|e_1,g_2\rangle+|g_1,e_2\rangle\big)$.
As the dipole is larger by a factor $\sqrt{2}$ compared to the dipole for
an individual emitter [cf. Eq.~\eqref{eq:HIrewrite}], the respective
radiative decay rate is enhanced by a factor of 2.

Here, we consider another effect of cooperative emission which occurs
when the emitted photons are indistinguishable,
even if the emitters are not confined to small volumes.
This effect can be best understood by considering the detection of two
photons with wave vectors $\mathbf{k}_1$ and $\mathbf{k}_2$, respectively.
The corresponding second-order coherence is given by
\begin{align}
  g^{(2)}_{\mathbf{k_1}\mathbf{k_2}}(t,\tau)=
  \frac{ \langle \sigma^+_{\mathbf{k}_1}(t)\sigma^+_{\mathbf{k}_2}(t+\tau)
  \sigma^-_{\mathbf{k}_2}(t+\tau)\sigma^-_{\mathbf{k}_1}(t)\rangle}
  {\langle \sigma^+_{\mathbf{k}_1}(t)\sigma^-_{\mathbf{k}_1}(t)\rangle
  \langle \sigma^+_{\mathbf{k}_2}(t+\tau)\sigma^-_{\mathbf{k}_2}(t+\tau)\rangle}.
\end{align}

With $\sigma^-_{\mathbf{k}_2}\sigma^-_{\mathbf{k}_1}
  =\cos\big((\mathbf{k}_2-\mathbf{k_1})\cdot\mathbf{r}/2\big)
  \sigma^-_1 \sigma^-_2$
and under the assumption of initially uncorrelated emitters with equal
excited state occupations, one finds

\begin{align}
  g^{(2)}_{\mathbf{k_1}\mathbf{k_2}}(t,0)
  %=\cos^2\big((\mathbf{k}_2-\mathbf{k_1})\cdot\mathbf{r}/2\big)
  =\frac 12\Big[1+\cos\big((\mathbf{k}_2-\mathbf{k_1})\cdot\mathbf{r}\big)\Big].
\end{align}

Thus, the measurement of a photon in mode $\mathbf{k}_1$ to which both QDs
are coupled collapses the state of the system onto an entangled state $|\psi_{\mathbf{k}_1}\rangle$.
In contrast to emission from uncorrelated states, the radiation pattern
for the emission from this entangled state is no longer uniform and
and emission into certain directions $\mathbf{k}_2$ becomes more likely
at the expense of others. This leads to an increased likelihood of measuring
two photons emitted into the same direction $\mathbf{k}_2=\mathbf{k}_1$ compared the
the case of independent emitters, although the overall radiative decay rate
does not experience a superradiant enhancement.

The predicted signatures of cooperative emission for our setup are summarised
in Figure~\ref{fig:1}(d) and \ref{fig:supmat_theory1}. The time dependence of $g^{(2)}$ is calculated by the minimal model
described in Section S1 under the assumption of incoherent pumping.
In the absence of cooperative emission (independent emitters),
one expects a dip in $g^{(2)}(\tau)$ with a value of $g^{(2)}(0)$
strictly limited to 0.5.
Any value larger than 0.5 therefore indicates cooperative emission
involving entanglement between the two emitters.

Both cooperative effects, superradiance and the
correlation-induced modification of the radiation pattern, lead to an
anti-dip in $g^{(2)}(\tau)$ with a value of $g^{(2)}(0)$ which is generally
unbound but $g^{(2)}(0)=1$ if the emission starts from
an initially uncorrelated state with equal excited state populations in each QD.
In the latter case, the width of the anti-dip is determined by local
dephasing with rate $\gamma_D$, while in the superradiant case the
incommensurablility between incoherent pumping and superradiant
decay reduces $g^{(2)}(\tau)$ for intermediate $\tau$ even in absence of
additional dephasing.

As $g^{(2)}(\tau)$ predicted in both types of cooperative effects are
qualitatively similar, superradiance and
the correlation-induced modification of the radiation pattern requires
can be distinguished by additionally measuring the radiative decay rate.
}

\section{Cooperative emission under continuous wave coherent driving: power dependence}\label{appendix:cw_cooperative}

\begin{figure*}
  \centering
  \includegraphics[width=0.7\textwidth]{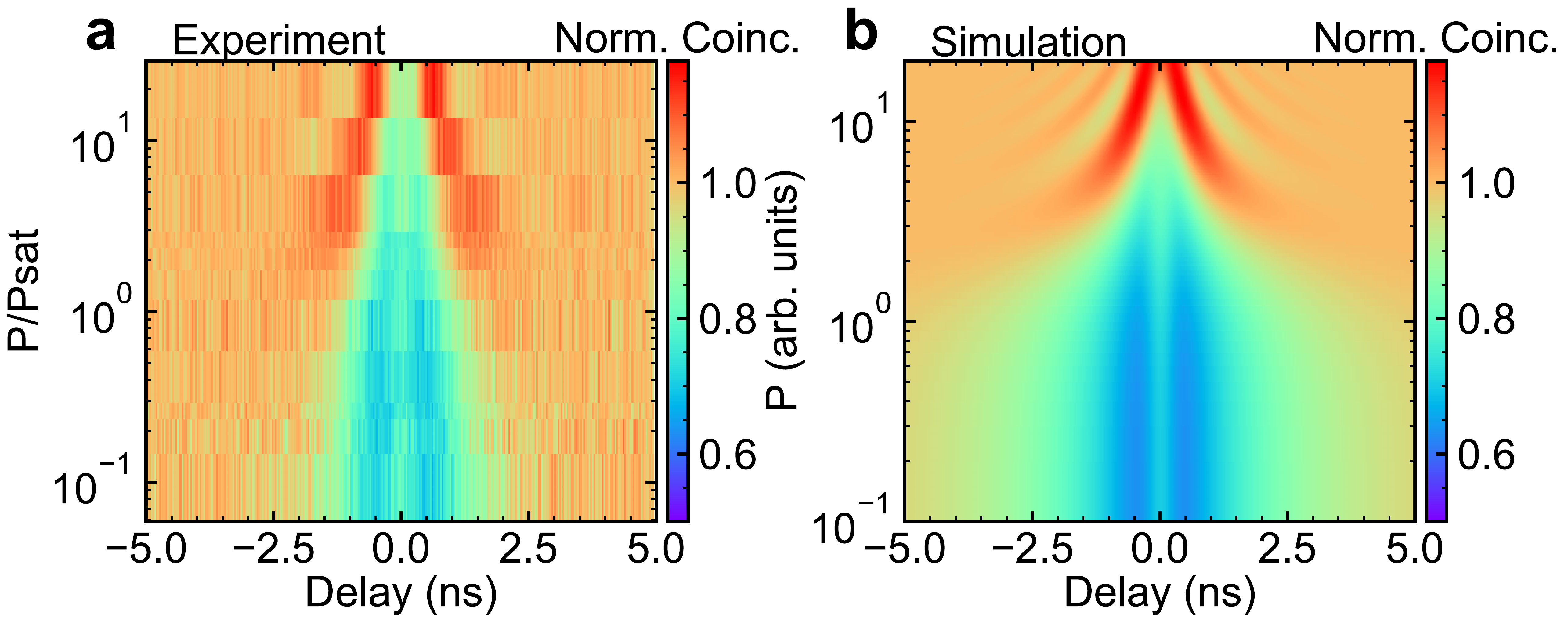}
  \caption{\textbf{Power dependence of the continuous wave resonant excitation of two QDs tuned into resonance.} A comparison of the experimental \textbf{a} and simulated data \textbf{b} for $g^{(2)}(\tau)$ as a function the ratio of the excitation power, P. The excitation power in the experimental data is normalized by the saturation power, Psat.}
  \label{fig:sr_CW_sweep}
\end{figure*}

Figure~\ref{fig:sr_CW_sweep} shows the theory-experiment comparison of the power dependence of the $g^{(2)}(\tau)$ data under continuous wave coherent (resonant) excitation of two indistinguishable quantum dots.
We observe Rabi oscillations as we increase the driving strength beyond the saturation power, Psat.
The experimental data agrees qualitatively with the simulated data obtained using the theoretical model defined in Section~\ref{appendix:minimalmodel} and the emitter parameters (e.g. spontaneous emission rate and dephasing rate), extracted from the experimental data in Figure~2d in the main manuscript.

\section{Cooperative emission under pulsed coherent driving: power dependence}\label{appendix:pulsed_cooperative}

\begin{figure*}
  \centering
  \includegraphics[width=1\textwidth]{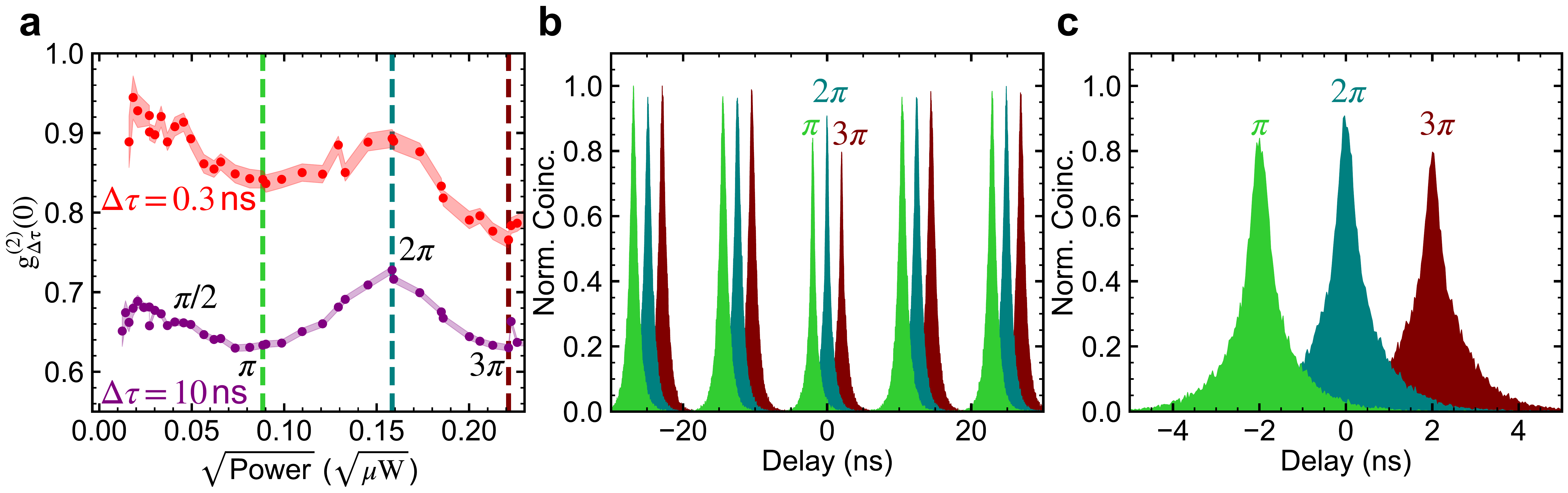}
  \caption{\textbf{Power dependence of the pulsed resonant excitation of two QDs tuned into resonance.}
    \textbf{a} Normalized zero-delay coincidences from the HBT measurement, $g^{(2)}_{\Delta \tau}(0)$ under resonant excitation, as a function of excitation power, within 0.3~ns (red) and 10~ns (purple) integration windows.
    \textbf{b, c} Coincidence histogram at a pulse area of $\pi$, $2\pi$ and $3\pi$ within 60 (\textbf{b}) and 10~ns (\textbf{c}) coincidence window. The data for $\pi$ and $3\pi$ are shifted temporally by -2 and 2~ns, respectively.
  }
  \label{fig:sr_pulsed_sweep}
\end{figure*}

Figure~\ref{fig:sr_pulsed_sweep} shows the power dependence of the $g^{(2)}(\tau)$ data under pulsed coherent (resonant) excitation of two indistinguishable quantum dots.
We demonstrate coherent control and the ability to manipulate the populations on the Dicke ladder as we vary the excitation power of the resonant pulse and observe that the $g^{(2)}_{\Delta \tau = 0.3\,\mathrm{ns}}(0)$ ($g^{(2)}_{\Delta \tau = 10\,\mathrm{ns}}(0)$) value oscillates with pulse area: a maximum of 0.90 (0.73) at $2\pi$ pulse area and a value of 0.84 (0.63) at $\pi$ pulse area.
The power dependence and the coincidence histograms that correspond to the three example pulse areas are depicted in Figure.~\ref{fig:sr_pulsed_sweep}(a) and ~\ref{fig:sr_pulsed_sweep}(b, c), respectively.
This result suggests that the measured $g^{(2)}_{\Delta \tau}(0)$ depends on the population of the bright, symmetric Dicke state: a minimal $g^{(2)}_{\Delta \tau}(0)$ value at pulse areas of odd multiples of $\pi$, where the system remains approximately in the ground state and a maximal $g^{(2)}_{\Delta \tau}(0)$ value at pulse areas of even multiples of $\pi$, where the system is close to the doubly-excited state which then collapses onto the symmetric Dicke state following the first photon detection event.
The contribution from the detection of any additional unintended photon, e.g. due to stray light or an increase of populations in the bright, symmetric Dicke state due to re-excitation during the pulse result in a increase in correlation and a hence a higher $g^{(2)}_{\Delta \tau}(0)$.
Conversely, the possibility of the re-excitation of the doubly-excited state along with the subsequent decay to the ground state during the remainder of the pulse result to a lower $g^{(2)}_{\Delta \tau}(0)$.
Full interpretation and modeling of this trend require the sufficient knowledge of multiple key parameters, including, but not limited to re-excitation rate~[48], excitation-induced dephasing rate~[45], and signal-to-background ratio due to laser leakage as a function of driving power.
%This complete model is beyond the scope of this manuscript~\cite{theory_paper}.

\section{Cooperative emission under pulsed incoherent driving} \label{appendix:nr}

\begin{figure*}
  \centering
  \includegraphics[width=0.65\textwidth]{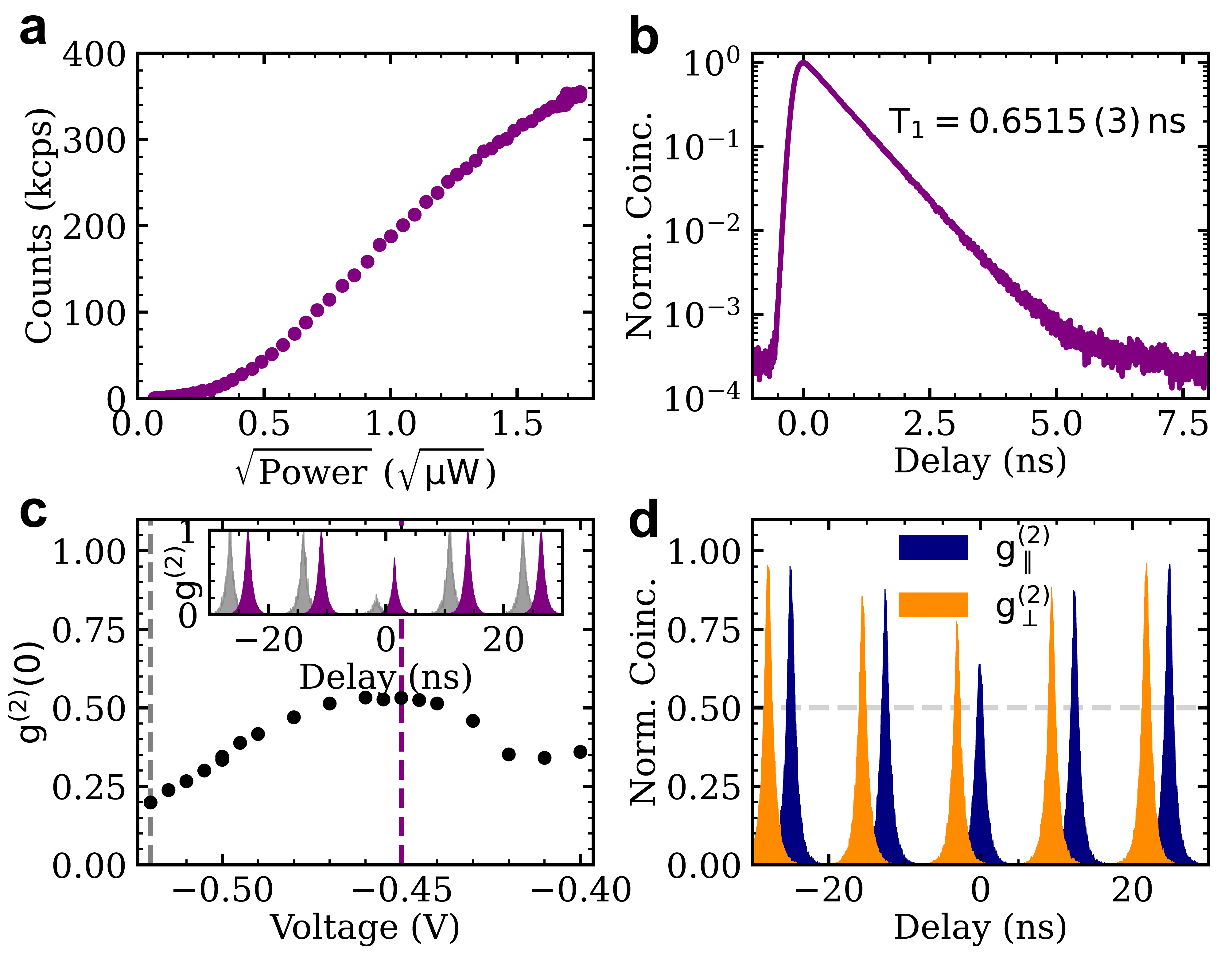}
  \caption{
    \textbf{Pulsed non-resonant excitation of two QDs tuned into resonance.}
    \textbf{(a)} Emission count rates from two QDs tuned into resonance, as a function of excitation power at bias of -0.45~V.
    \textbf{(b)} Time-resolved emission profile of the cooperative emission at power of $0.58\,\mu\mathrm{W}$.
    \textbf{(c)} Normalized zero-delay coincidences from the HBT measurement, $g^{(2)}(0)$ as a function of applied bias (excitation power = $0.58\,\mu\mathrm{W}$).
    Inset shows the normalized coincidences from the HBT measurement at -0.52~V ($\delta\approx 95\,\mu\mathrm{eV}$, grey) and -0.45~V ($\delta \approx 0\,\mu\mathrm{eV}$, purple).
    \textbf{(d)} Two-photon interference between subsequently emitted photons, separated by 12.44~ns, for the case where input photons, from both QDs tuned into the resonance ($\delta=0$) are prepared in parallel ($g^{(2)}_\parallel$, blue) and perpendicular ($g^{(2)}_\perp$, orange) polarization. The non-post-selected (post-select within 100~ps window) interference visibility for this measurement is 0.041 (0.212).
    All measurements are done with 14~ps excitation pulses, frequency detuned by $\approx 0.4~\mathrm{meV}$ from the two-dot-degeneracy at $\lambda = 971.169\,\mathrm{nm}$.
  }
  \label{fig:supmatnr2}
\end{figure*}

Please refer to the captions of Figure~\ref{fig:supmatnr2} for additional data on the spectroscopy of the emission from two QDs under pulsed incoherent, non-resonant (phonon-assisted) excitation.

\section{Correction to the degree of entanglement due to noise and detection jitter}\label{appendix:entanglement_fidelity}

In this section, we provide an analysis of the degree of entanglement for simultaneous detection of the emission from two indistinguishable emitters in the presence of the noise and detection jitter.
As described in Section~8 of the Supplementary Material in Ref.~[10], one can extract the ratio of noisy photons compared to single photons $p_{n}$ extracted from the $g^{(2)}_\mathrm{QD1}(0)$ for individual QD emission, i.e.
\begin{equation}
  g^{(2)}_\mathrm{QD1}(0) = \frac{p_n(2+p_n)}{(1+p_n)^2}.
\end{equation}
Using this information, we can obtain the degree of entanglement, given by the fidelity to the maximally entangled symmetric Dicke state $(1/\sqrt{2}\left(\ket{e_1,g_2}+\ket{g_1,e_2}\right))$, $\mathcal{F}$ using
\begin{equation}
  \mathcal{F}\geq \frac{(g^{(2)}(0)(1+p_n)^2-p_n(2+p_n))^2}{g^{(2)}(0)(1+p_n)^2}.
  % g^{(2)}_\mathrm{Corr.}(0)\geq \frac{(g^{(2)}(0)(1+p_n)^2-p_n(2+p_n))^2}{g^{(2)}(0)(1+p_n)^2}.
\end{equation}
% where $g^{(2)}(0)$ is the normalized (raw) zero-delay coincidence for two indistinguishable emitters.

\begin{figure*}
  \centering
  \includegraphics[width=\textwidth]{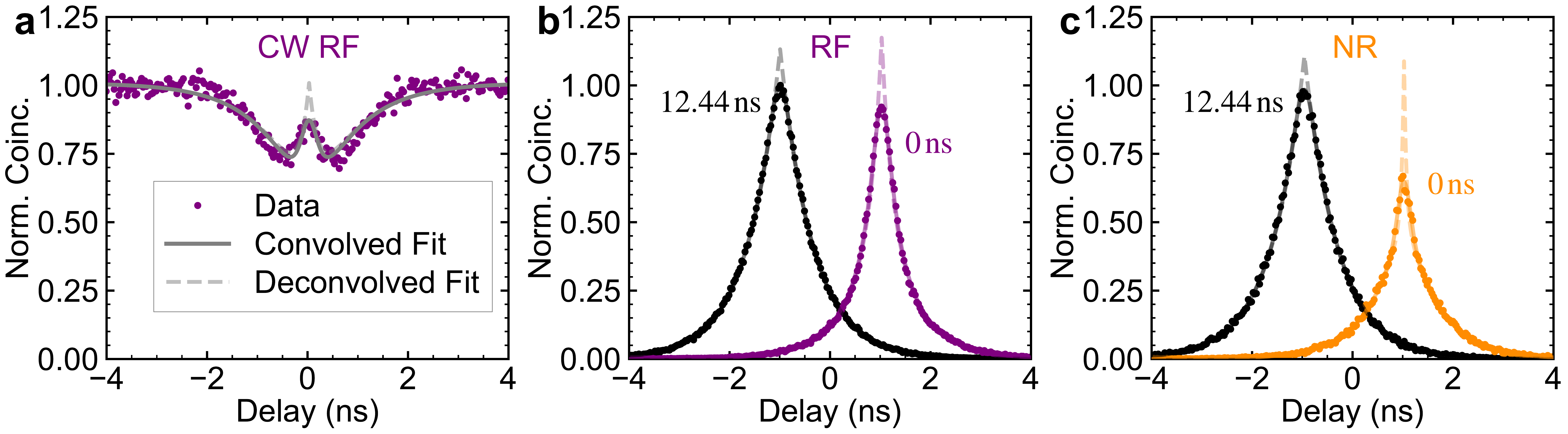}
  \caption{\textbf{Comparison between experimental data, the convolved and deconvolved fit for the second order intensity correlation measurement with two indistinguishable emitters.}
    Normalized coincidence (to the raw coincidences at longer delays of $\approx 10\,\mu\mathrm{s}$) under continuous wave resonant (CW RF, \textbf{a}), pulsed resonant (RF, \textbf{b}) and non-resonant excitation (NR, \textbf{c}).
    Convolved and deconvolved fits are given by the solid and dashed lines, respectively.
    Experimental data and the convolved fits are the same as in Figure~2d (CW) and 3d (Pulsed).
    The uncorrelated side peak at 12.44~ns and the correlated zero-delay peak are shifted by -1 and 1~ns, respectively.
  }
  \label{fig:S4_deconv}
\end{figure*}

In the noiseless limit (i.e. $p_n=0$), this is given solely by the measured (normalized) coincidences at the zero-delay in the second-order intensity correlation function $g^{(2)}(0)$.
The presence of surplus ``noisy" photons in the signal (which may originate from further emitters or the excitation laser) and the timing jitter in the detection setup result in a decrease in the entanglement fidelity $\mathcal{F}$.
To account for both effects, we extract the ratio of noisy photons in the single photon stream from an intensity correlation measurement with only a single emitter.
We fit the experimental data in Figure~2d and 3d by convolving the theoretical expressions in Eq.~1 and 2 with the independently measured detection timing jitter (full-width-at-half-maximum of $0.240\,\mathrm{ns}$) and compare them with the corresponding deconvolved fits.
The fits, along with the experimental data are depicted in Figure~\ref{fig:S4_deconv}. We summarise the results, i.e. the change in the $g^{(2)}(0)$ value with respect to each imperfections in Table~\ref{tab:figure_of_merit}.

It is important to note that after deconvolution, the zero-delay peak (pulsed excitation) and the anti-dip (continuous wave excitation) have a maximum close to one, as expected from the ideal noise-less case for the emission from two indistinguishable emitters.
This shows that the maximum of the zero-delay peak is largely determined by detector jitter and hence is inadequate to quantify the entanglement of the emitters.
This is analogous to the interpretation of the zero-delay anti-bunching dip in the Hong-Ou-Mandel interference measurement of subsequently emitted single photons; please refer to Refs.~[42-44] for more details.
Instead, the integration over the full zero-delay peak within a large integration window ($g^{(2)}_{\Delta \tau}$) gives information about the dephasing of the emitted photons, independent of the detector jitter, and hence is more suitable and useful to characterise the entanglement of the emitters.

\begin{threeparttable}[!ht]
  \caption{The degree of entanglement, indicated by the normalized zero-delay coincidences $g^{(2)}(0)$ from two degenerate QDs under CW and pulsed resonant (RF) and non-resonant (NR) excitation in the presence of noise, given by the ratio of noisy photons in the QD signal $p_n$ and the detection timing jitter (modelled as a Gaussian with a full-width-at-half-maximum of $0.240\,\mathrm{ns}$).
  The correction to this quantity, given by the fidelity to symmetric Dicke state $\mathcal{F}$, as a result of the presence of noisy photons is estimated using the expression in Sec.~8 of the Supplementary Material in Ref.~[10] using the extracted ratio of noisy photons to QD signal, $p_n$ in the single emitter case.
  For the case with pulsed measurements, the values reported here are extracted from the coincidence histogram with integration windows of 10~ns ($g^{(2)}_{\Delta \tau=10\,\mathrm{ns}}(0)$).
  The values with temporal post-selection of $\leq 0.1$~ns ($g^{(2)}_{\Delta \tau\leq 0.1\,\mathrm{ns}}(0)$) are listed in parenthesis.}
  \label{tab:figure_of_merit}
  \begin{ruledtabular}
    \begin{tabular}{|l||c|c|c|}
                                                & \multicolumn{1}{c|}{CW}      & \multicolumn{2}{c|}{Pulsed}                   \\
      \hline
      \textrm{}                                 & \hspace{.5cm}\textrm{RF$^a$} &
      \textrm{RF$^b$}                           &
      \textrm{NR$^c$}                                                                                                          \\
      \hline
      \colrule
      Measured QD1 $g^{(2)}_\mathrm{QD1}(0)$    & 0.06                         & 0.05 ({  0.05})             & -               \\ [.2cm]
      Ratio of noisy photons $p_n$              & 0.031                        & 0.026 ({  0.026})           & -               \\ [.2cm]
      Measured QD1+QD2 $g^{(2)}(0)$             & 0.87                         & 0.67 ({  0.93})             & 0.56 ({  0.65}) \\ [.2cm]
      Deconvolved $g^{(2)}_\mathrm{Deconv.}(0)$ & 1                            & 0.68 ({  1.04})             & 0.54 ({  0.98}) \\[.2cm]
      $\mathcal{F}$ $^d$                        & 0.80                         & 0.60 ({  0.88})             & 0.56 ({  0.65}) \\[.2cm]
      $\mathcal{F}^{\mathrm{Deconv.}}$ $^e$     & 0.94                         & 0.61 ({  0.99})             & 0.54 ({  0.98}) \\ [.2cm]
    \end{tabular}
    \begin{tablenotes}
      \item[a] Data taken at $\mathrm{P}\sim 0.3\,\mathrm{P_{sat}}$
      \item[b] Data taken at $\pi/2$-pulse
      \item[c] Data taken at saturation.
      \item[d] Correction to the $g^{(2)}(0)$.
      \item[e] Correction to the $g^{(2)}_\mathrm{Deconv.}(0)$.
    \end{tablenotes}
  \end{ruledtabular}
\end{threeparttable}

%\bibliography{reference}